\documentclass [aps,preprint,showpacs]{revtex4}
\usepackage{graphicx}
\begin{document}
\title{\bfseries Symmetry interaction and many-body correlations}
\author {M. Papa$^{1}$ and G. Giuliani$^{1,2}$ \\
$^{1}$Istituto Nazionale Fisica Nucleare-Sezione di Catania \\
Via S. Sofia 64 Catania 95123 Catania, Italy \\
$^{2}$Dipartimento di Fisica ed Astronomia, Università di Catania \\
Via S. Sofia 64 Catania 95123 Catania, Italy \\
}
\linespread{1.2}
\setlength{\textwidth}{16cm}
\setlength{\oddsidemargin}{-0.5mm}

\date{}

%

\begin{abstract}{ It is shown how many-body correlations involving the
symmetry potential naturally arise in the molecular dynamics CoMD-II
model. The effect of these correlations on the collision dynamics at
the Fermi energies is discussed. Small level of correlations for
systems of moderate asymmetry are able to produce large effects also
in simple observables like the charge distributions. The comparison
with predictions based on EOS static calculations is also
discussed.}
\end{abstract} \pacs{25.70.Pq, 02.70.Ns, 21.30.Fe, 24.10.Cn}
\maketitle
\section{Introduction}
Great efforts have been made in the last decades in trying to
extract information on the behavior of the symmetry energy as a
function of the density.  Semiclassical microscopic
approaches predict that the
 isotopic composition of cluster productions, at Fermi or higher energies, induced by
nucleus-nucleus collisions lead to the definition of well suited
observables able to highlight the degree of "stiffness" of the
symmetry interaction \cite{epj}. A wide class of phenomenological
effective two and three-body interactions are used in semi-classical
calculations (see for examples the class of Skyrme interactions
\cite{rikowsca} and Gogny forces) for which  the density of symmetry
interaction is usually expressed as: $\varepsilon^{\tau}=e_{a}\rho
F(\rho/\rho_{0})\beta^{2}$
(the $\beta^{4}$
dependence is usually considered negligible).
$\beta=\frac{\rho_{n}-\rho_{p}}{\rho}$ where $\rho_{n}$, $\rho_{p}$
and $\rho$ are the neutron, proton and total densities respectively.
$\rho_{0}$ is the groud state (g.s.) density.
 These results are consistent with in medium EOS
calculations using microscopic effective interactions (see as a
typical example ref.\cite{bao0,lattimer,fucks}).  At density lower
than 1-1.5 $\rho_{0}$, $e_{a}F$ is usually a positive monotonic
increasing function of the density, i.e. it produces in most of the
cases a repulsive behavior as a function of the density. In
semiclassical transport models the above expressions are used
locally to generate the single particle potentials (see as an
example BUU calculations \cite{bao0}).

The calculations based on realistic effective interactions using
variational techniques  and cluster expansion methods start from
quite complex two and three body interactions. As suggested also
from Energy Density Functional theories \cite{edf},
 the one-body total
density functional can represent the effect of rather complex
interactions and correlations quite difficult to be managed in time
dependent approaches. However, we note that the presence of the
$\beta^{2}$ variable, as the leading factor determining the
iso-vectorial dynamics (see as an example
ref.\cite{lattimer,fucks}), does not take into account correlations
able to describe the different dynamics between neutron-neutron
($nn$), proton-proton ($pp$) and neutron-proton ($np$) couples.

In fact, it is easy to verify that the $\beta^{2}$ dependence
implies that the probability to find a $np$ couple in a small volume
V (local description) is just the product of the probability to find
a $pp$ and a $nn$ couple in the same volume. With this respect the
$\beta^{2}$ dependence of the symmetry interaction resembles a kind
of mean-field approximation concerning the two-body iso-vectorial
dynamics (which in the following we indicate with I.M.F.A.). At the
Fermi energies, however, many-body correlations involving also the
isospin degree of freedom are necessary to describe the cluster
formation process; a closer investigation on the behavior of the
effective interaction including such correlations in a time
dependent picture would be desirable \cite{arch}. In this energy
range, in fact, the formation of clusters and the related dependence
on the symmetry potential, are the results of the many-body fast
evolution leading the investigated systems in regions of density and
excitation far from the $g.s.$ one.
 The study of observables having a strict link with this fast changes
can give the opportunity to probe the density functional dependence
of the microscopic interaction. In sect II  we discuss some details
concerning the symmetry interaction as treated in our Constrained
Molecular Dynamics Model-II (CoMD-II)\cite{comd1,comd2}. In section
2, as an example,
 we illustrate results of the performed calculations for the
$^{40}Cl+^{28}Si$ at 40 MeV/nucleon. The effects of correlations on
the two-body iso-vectorial dynamics are also discussed. Section 3
contains the conclusive remarks.

\section{Symmetry energy and the CoMD-II model}

As it is well known the symmetry energy contribution to the total
energy include a kinetic term and a potential one. In CoMD-II model
the kinetic part is naturally included through the constraint
related to the Pauli principle acting on the phase space of the
neutron and proton fluids. The increasing of the kinetic term with
the increasing of the charge/mass asymmetry is therefore included in
our calculations. In particular in ref.\cite{jcomp} it is shown how
the constraint is able to produce for finite systems  Fermi like
distribution during all the dynamical evolution of the investigated
system.

In the following we instead describe how the symmetry interaction is
included in the model. From a general microscopic point of view the
symmetry potential can be introduced by assuming that, in $S$ waves
(see the following), the two body interaction for triplet isospin
states (T=1) $\Upsilon^{1}$ is repulsive as compared to the more
attractive singlet states (T=0) $\Upsilon^{0}$ i.e.
$\Upsilon^{1}>\Upsilon^{0}$ (both $\Upsilon^{1}$ and $\Upsilon^{0}$
are negative) . These are well known facts established from low
energy nucleon-nucleon scattering processes, from studies concerning
the binding energy of the deuteron and the ones related to binding
energies of isobars nuclei.

According to the properties of  the many-body wave function
describing the system,  the triplet interaction is experienced by
 $(N^{2}+Z^{2}+NZ)/2$ couple of states, while the singlet one involves
 $NZ/2$ couples. $N$ and $Z$ indicate the number
of neutrons and protons respectively for a system with mass number
$A$. By using a Skyrme interaction framework, the above
considerations are equivalent to use a symmetry microscopic
interaction having the following structure:
\begin{eqnarray}
V^{\tau} &=& \frac{a_{0}}{2\rho_{0}}\sum_{i\neq
j=1}^{A}(2\delta_{\tau_{i}\tau_{j}}-1)
  \delta(\overrightarrow{r}_{i}-\overrightarrow{r}_{j})
\end{eqnarray}
with $a_{0}=(\Upsilon^{1}-\Upsilon^{0})/2=72 MeV$. $\rho_{0}$ is the
ground state (g.s.) density.
 $\overrightarrow{r}_{i}$ indicates the
generic nucleonic coordinate. The above assumptions generate also an
attractive iso-scalar interaction $V$ related to the contribution
$A(3/8\Upsilon^{1}+1/8\Upsilon^{0})$ \cite{comd1,comd2}.

By performing the convolution of eq.(1) with the nucleonic wave
packets,  we  obtain the following expression for the total symmetry
energy:
\begin{eqnarray}
 U^{\tau} &=& \frac{a_{0}}{2\rho_{0}}\beta_{M} \\
 \beta_{M}  &=& \rho^{nn}+\rho^{pp}-2\rho^{np}\\
\rho^{aa'} &=& \sum_{i\neq j}^{i\subseteq a,j\subset a'}\rho_{i,j}
\end{eqnarray}

with $a$ and $a'$ equal to $n$ or $p$ being $n$ and $p$ the ensemble
of neutrons and protons respectively.
 $\rho_{i,j}$ represents the normalized two-body Gaussian
overlap integral  typical of  molecular dynamics approaches. The
above expressions represent the contribution related to the leading
two-body contribution to the symmetry interaction. We note that, due
to the character of our molecular dynamics approach, the elementary
quantities appearing in the above expressions are the overlap
integrals between the different couples.  Usually, as suggested from
realistic in medium variational calculations the above expression is
implemented by introducing form factors which are able to take into
account the effects of more complex in medium interactions quite
difficult to be managed in time dependent approaches. In all the
time dependent semiclassical approaches \cite{epj}, however, this
form factors are introduced as phenomenological correction factors
whose parameters should be definitely obtained through the
comparison with the experimental data \cite{epj}.

For clarity, in the following we report the expressions obtained in
the so called Non-Local (N.L.) approximation which is essentially
valid for non light systems in compact configurations. The following
expressions are in fact more simple  than the ones used in the
calculations being able to easily enlighten  the roles played by the
leading factor $\beta_{M}$  and by the form factors.
\begin{eqnarray}
U_{N.L.}^{\tau} &=& \frac{ a_{0} }{2\rho_{0}}F'(s)\beta_{M} \\
F'(s) &=& \frac{2s}{s_{g.s.}+s}\;\;\;\;\;\;\;\;\;\; Stiff1 \\
F'(s) &=& 1\;\;\;\;\;\;\;\;\;\;\;\;\;\;\; \;\;\;\;\;Stiff2 \\
F'(s) &=&(\frac{s}{s_{g.s.}})^{-\frac{1}{2}}\;\;\;\;\;\;\;\; Soft \\
s &=&\frac{4}{3A} \sum_{i \neq j}^{i,j\subseteq a\cup a'}\rho_{i,j}
\end{eqnarray}
In particular, the N.L. approximation is easily obtained by
substituting the generic overlap integral $s_{j}=\frac{\sum_{i\neq
j}^{i \subseteq a\cup a'}\rho_{i,j}}{2A}$ for the j-th particle
appearing in the complete expression, with the average value related
to the $A$ particles of the system $s_{j}\simeq s/2A$. $s$ according
to eq. (9) is associated to the total overlap integral.

 The different options concerning the form factors describe
different degrees of stiffness. For the system under study,
$^{40}Cl+^{28}Si$ at 40 MeV/nucleon, and for compact configurations,
$s$ is well approximated by the  one-body density $\rho$.
 This value produces a total strength factor which  is in substantial agreement with the ones
 obtained by EOS in medium
 calculations (see the following).   We note that
the positive form factors $F'(s)$ are such that
$F'(s)\frac{s}{s_{g.s.}}$ has the same functional form like the
$F(u)$ proposed in \cite{bao0,wiringa,lattimer}. They have been used
extensively in BUU calculations.  The factor $\beta_{M}$ already
introduced in eq. (3), instead arises naturally from the many-body
approach. It takes into account explicitly, through the last
negative term, that the microscopic two-body interaction in the
isospin singlet states is more attractive than the one related to
triplet states.

For consistency and clarity, before showing the results of our
calculations, we want now evaluate the previous expressions in
I.M.F.A.. For this purpose we assume, for simplicity, $A,N,Z\gg 1$,
and we rewrite $\beta_{M}$ as:
\begin{eqnarray}
\beta_{M}& = & N^{2}\tilde{\rho}^{nn}+
Z^{2}\tilde{\rho}^{pp}-2NZ\tilde{\rho}^{np}.
\end{eqnarray}
where: $\tilde{\rho}^{aa'}=\rho^{aa'}/II'$ with $a$ and $a'$ equal
to $n$ or $p$ and correspondingly $I$ and $I'$ equal to $N$ or $Z$.
$\tilde{\rho}^{aa'}$ represents the average overlap integral per
couple of nucleons.

\textit{I.M.F.A. is now easily obtained by supposing
$\tilde{\rho}^{aa'}=\frac{\sum_{i\neq j=1}^{A}\rho_{i,j}}{A^{2}}
\equiv\tilde{\rho}=\frac{3}{4A}\rho$ to be independent on $a$ and
$a'$}. This is a rather strong approximation in our framework, and
it is equivalent to assume same average distances $d_{a,a'}$ between
nucleons in $nn$, $pp$ and $np$ couples. This means that
correlations producing  differences in the two-body iso-vectorial
dynamics $\tilde{\rho}^{aa'}$ are averaged through the overlap
integral $\tilde{\rho}$, so that:
$\beta_{M}\rightarrow\frac{3}{4}\frac{\rho}{A}(N-Z)^{2}=\frac{3}{4}\rho^{2}V
\frac{(\rho_{n}-\rho_{p})^{2}}{\rho^{2}}$. In this case we obtain
that the interaction depends locally only through one-body
densities, it is positively defined and the $\beta^{2}$ dependence
is restored. Taking into account  that for the present system, in
compact configurations, $s\cong\rho$, by substituting the right hand
of the above relation in eq. (4) and dividing by the volume $V$, we
obtain a precise correspondence with the symmetry potential density
$\varepsilon^{\tau}$ used in ref.\cite{bao0} (see also section 1) if
$e_{a}=27 MeV$. This corresponds  to an $S_{0}$ value \cite{bao0} of
about 40 MeV.

So that, unambiguously,  the microscopic interaction defined through
eq. (1) is able to generate a symmetry term strictly equivalent, in
the spirit of the I.M.F.A., to the expression obtained through EOS
calculations like the one reported in ref.\cite{lattimer}.

\section{Results of  calculations}
In this section, we illustrate the results obtained with our model
for the system $^{40}Cl+^{28}Si$ at 40 MeV/nucleon.

For moderately asymmetric systems, self-consistent calculations,
including the ones for the searching of the g.s. configurations,
produce a negative value of $\beta_{M}$ (see eq. (3)). This result
is due to the combined actions of the more attractive force, for the
isospin singlet states (which is experienced only for $np$ couples),
and the repulsive effect of the Coulomb interaction for $pp$
couples. The Pauli principle also plays its role in the symmetry
interaction. In our model calculations it becomes dominant for light
systems.
 These correlations tend to increase the average neutron-proton overlap
integral $\tilde{\rho}^{np}$ (decreasing the related average
distance $d_{np}$) and to decrease the overlap $\tilde{\rho}^{nn}$
and $\tilde{\rho}^{pp}$ related to the $nn$ and $pp$ couples
(deuteron effects). As an example, for the system under study and
for an impact parameter b=4 $fm$ , in Fig. 1(a) we show, as a
function of time, the ensemble average overlap integral
$\overline{s}$ for the three investigated cases ($Stiff1$,$Stiff2$,
$Soft$) related to the total system.  During the time interval 18-60
fm/c about 95\% of the total system is formed by a compact cluster
having, for the different options, the same mass and the same
charge/mass asymmetry within 1\%. In this way, the hot compact
system, in all the three cases, experiences almost adiabatic
compression and decompression processes from which the density
dependence of the symmetry interaction stored in the system can be
investigated.

\begin{figure}
  \includegraphics[height=7.5cm, width=7.5cm, angle=0,  keepaspectratio]{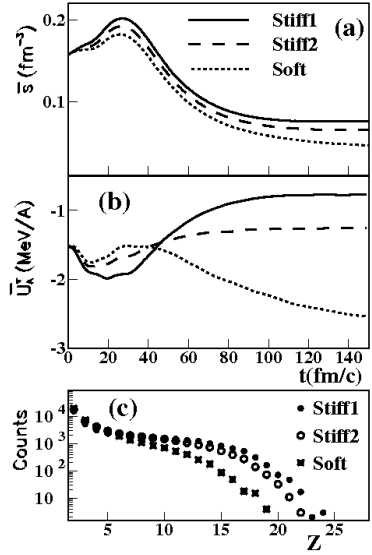}
\caption{For the system $^{40}Cl+^{28}Si$ at 40 MeV/nucleon and
  b=4 fm, we show:
(a) average overlap integral $\overline{s}$, (b) average symmetry
potential per nucleon as a function of time for the different
options describing the symmetry potential (see eqs.(6-8)), (c)
charge distribution for $Z>1$ obtained after 650 fm/c. }
\end{figure}

From Fig. 1(a) we can see that the maximum value of $\overline{s}$,
$s_{max}$, is reached in about 30 fm/c. The $Soft$ case shows the
lower $s_{max}$ value, the $Stiff1$ the higher one (about 1.25 times
the $s_{g.s.}$ value). Accordingly, the average total potential (not
shown) displays the less attractive behavior  for the $Soft$ case.
The reasons of these differences can be understood by looking at
Fig. 1(b) where we show the corresponding average symmetry potential
per nucleon $\overline{U}_{A}^{\tau}$. $\overline{U}_{A}^{\tau}$
shows, in fact, an average repulsive effect as a function of the
density for the $Soft$ case (it increases with the density), while
the $Stiff1$ case gives rise to an attractive effect.

 The $Stiff2$ case shows  an attractive behavior and
intermediate values concerning the strength. The  differences
observed around 18-60 fm/$c$ play an important  role in to determine
the later evolution of the system. In Fig.1(c) we show the charge
distributions evaluated for the three different options after 650
fm/c (main fragments almost cold). The rather high sensitivity of
this simple observable to the different shapes of the form factors
is clearly evident. In particular, in our calculations we have
verified that
 $\tilde{\rho}^{np}=\tilde{\rho}^{pn}\equiv(1+\alpha)\frac{(\tilde{\rho}^{pp}+
 \tilde{\rho}^{nn})}{2}\equiv\tilde{\rho}$
with $\alpha\simeq 0.15$. Therefore, at low asymmetry, a small
correlation effect is enough to produce a negative value of
$\beta_{M}$, due to the structure of eq. (2). Moreover the strength
of $\overline{U}_{A}^{\tau}$ is rather high, of the order of -1.5
MeV/nucleon for $t=0$ $fm/c$ at normal density.
Therefore, according to these results, we observe that also a
correlation value of the order of 1\%-2\% can induce non negligible
effects on the symmetry term. This strongly suggests that the
effects discussed in this section can be still present by using
other kinds of approaches and effective interactions and that they
refer to a rather general aspect of the symmetry term  due to its
intrinsic structure.

Concerning the negative sign, we would like to precise  that
$\beta_{M}$ shows always a repulsive character with the asymmetry of
the system. This repulsive effect  follows an approximate parabolic
dependence with respect to the $\chi=N-Z$ variable. In particular:

\begin{eqnarray}
\beta_{M}& \sim & \tilde{\rho}\{\chi^{2}[1+\alpha/2]-\alpha
A^{2}/2\}
\end{eqnarray}
where the correlation coefficient $\alpha$ is a function of N,Z and
$\tilde{\rho}$. $\alpha$ has to be evaluated through self-consistent
calculations. This means that $\beta_{M}$ contains, independently on
its sign,
 the right behavior necessary to explain the differences in the
binding energies of isobars nuclei.
\begin{figure}
  \includegraphics[height=9.cm, width=9. cm, angle=0,
  keepaspectratio]{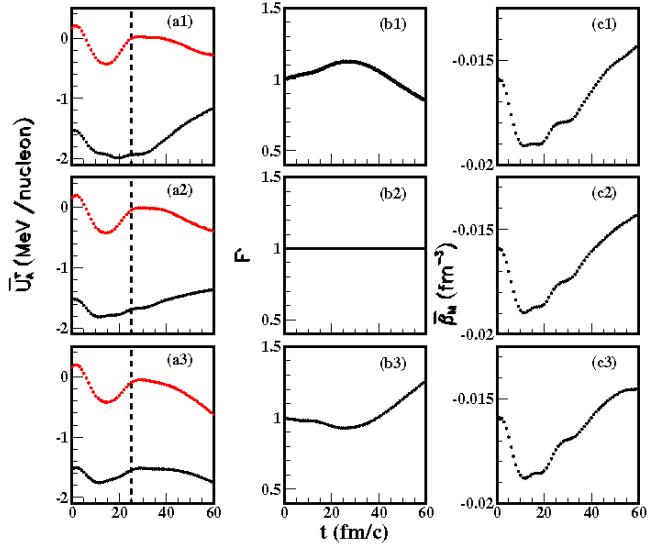}\\
  \caption{In the generic panels (ai),(bi),(ci) are reported the
  average symmetry potentials per nucleon $\overline{U}_{\textrm{A}}^{\tau}$, the
  form factors $F'$ and the leading average factor $\overline{\beta}_{\textrm{M}}$
  respectively. The index i=1,2,3 are referred to the different options for the
  symmetry potential Stiff1, Stiff2 and Soft respectively. The black symbols
  concern the system under study. The red symbols represent results for
  a system having the same total mass and 27 units of charge.}
\end{figure}
For systems with mass around 68 units, the $\beta_{M}$ factor
becomes positive for relevant "limiting" asymmetries
restoring therefore the repulsive behavior with the density of the
asy-stiff cases here investigated. However, we stress that the
negative sign of $\beta_{M}$ has  relevant consequences in to
determine the effects of the different options for $U^{\tau}$ on the
dynamics of the investigated processes. Contrary to the uncorrelated
case, the so called Stiff parameterizations (in spite of this change
of sign, we retain the same nomenclature to indicate the different
options) show an average decreasing behavior with the density. For
the system under study an overall description, as a function of
time, of the different factors which define the symmetry interaction
per nucleon is shown in fig.2.

With black symbols in the tree columns of panels (a), (b),(c) we
plot $\overline{U}_{\textrm{A}}^{\tau}$, the form factors $F'$ and
the leading factor $\overline{\beta_{\textrm{M}}}$ respectively. The
three rows are referred to the three options Stiff1, Stiff2, and
Soft respectively. The vertical dashed line enlighten the time
around which the average density reaches the maximum value. From the
figure it is possible to see how the interrelated behavior of the
factors $F'$ and $\beta_{\textrm{M}}$ determines the final evolution
of the symmetry potential for the different options. The red symbols
on the left column indicate the $\overline{U}_{\textrm{A}}^{\tau}$
values for a system having the same mass of the $^{40}Cl+^{28}Si$
system but a charge/mass asymmetry $\frac{N-Z}{A}$ of about 0.22.
These results have been obtained, as a first order approximation, by
supposing that the average overlap integral (per nucleon couples)
$\tilde{\rho}$  for this new system follows the same time behavior
as the one associated to the $^{40}Cl+^{28}Si$ system and applying
eq. (11) (full self-consistent calculations are in progress). These
estimations show clearly that for the stiff cases, around 25 fm/c,
$\overline{U}_{\textrm{A}}^{\tau}$ becomes close to zero and the
behavior changes from an attractive action to a repulsive one as a
function of the density. This change of behavior concerns only the
stiff cases analyzed in the present work and not the soft option
which produces always a repulsive action. This transition,
therefore, could be considered as a "fingerprint" of a finite value
for the correlation coefficient $\alpha$  in the case of a stiff
behavior of the symmetry interaction.

We conclude this section by showing in Fig.3 the same quantities as
plotted in Fig.1 but referred to our self consistent calculations
obtained with the I.M.F.A. discussed at the end of the previous
section. It is clearly evident that the further constraint
reflecting the absence of iso-vectorial two-body correlations
($\alpha=0$) has noticeable effects. We observe a quite reduced
strength of the symmetry interaction which now assume positive
values (with an order of magnitude valuable also in the framework of
the Liquid Drop Model) and a substantial repulsive effect for all
the three options. The final results on the charge distribution,
accordingly, show now an almost independence on the different
options.
\begin{figure}
  \includegraphics[height=7.5cm, width=6.5cm, angle=0., keepaspectratio]{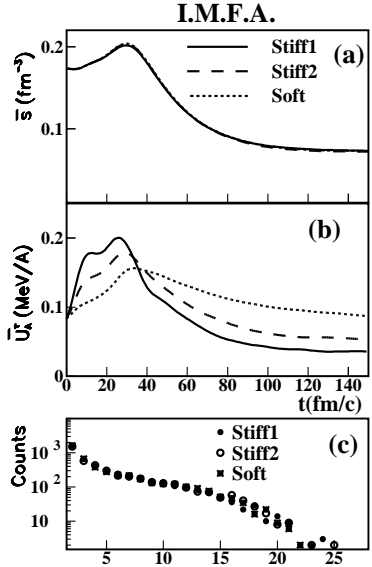}
  \caption{
Same quantities like the ones shown in Fig.1 are plotted in the case
of the so called Isovectorial Mean Filed Approximation (I.M.F.A) }
\end{figure}

\section{Conclusive remarks}
Non-vanishing correlations in the neutron-proton dynamics related to
the symmetry interactions, and obtained with the semiclassical
CoMD-II model, strongly affect the dynamics of heavy ion collisions
at the Fermi energies. These many-body correlations suggest that the
leading factor responsible for the microscopic two-body
iso-vectorial dynamics can have a structure  different from the
usual $\beta^{2}$ dependence suggested by  EOS calculations like the
ones reported in ref. \cite{lattimer}. At moderate asymmetries these
effects can produce noticeable change in the dynamics of the heavy
ion collisions. In particular, the Stiff options here investigated
show in the low density region (contrary to the I.M.F.A. case
discussed in this paper) an average attractive behavior as a
function of the density. Preliminary calculations also suggest that
the strength of this attractive effect decreases with the asymmetry
of the system and that, for values larger than some limiting value
$\beta_{c}$ ($|\beta_{c}|>0.22$), the repulsive behavior is
restored.

We think that the definitive answer about the real entity  of such
correlations and their change with the system asymmetry should be
obtained through further deep investigations on the behavior of
different isospin observables to be measured experimentally. This
subject can have a certain relevance in the isospin physics studied
through heavy ion collisions. In fact, it has been shown that time
dependent semiclassical approaches including such correlations can
give quite different results with respect the ones based on the
I.M.F.A. This, therefore, allows for different conclusion about the
estimation of the "stiffness" degree of the symmetry interaction
when comparing calculations with experimental data.
Finally we remark that, the large effects produced also through a
small degree of correlations, as predicted by our model
calculations, suggest that, in the quite sophisticated static EOS
calculations, the inclusion of variational parameters having a
charge/mass asymmetry dependence (as performed in studies related to
pairing forces \cite{shuck,edf}) should be investigated in more
detail.

\end{document}